\documentclass[%
reprint,
 amsmath,amssymb,
 aps,
]{revtex4-1}

\usepackage{graphicx}% Include figure files
\usepackage{dcolumn}% Align table columns on decimal point
\usepackage{bm}% bold math
\begin{document}

\preprint{APS/123-QED}

\title{A fractional generalized Cauchy process}% Force line breaks with \\

\author{Yusuke Uchiyama}
\author{Takanori Kadoya}%
\affiliation{%
MAZIN, Inc., Minami-Otsuka, Toshima, Tokyo 170-0005, Japan
}
\author{Hidetoshi Konno}%
\affiliation{%
Emeritus Professor, University of Tsukuba, Tsukuba, Ibaraki 305-8573, Japan
}%
%\affiliation{%


%\affiliation{
%\affiliation{
%\affiliation{%


\date{\today}% It is always \today, today,

\begin{abstract}
This paper presents a fractional generalized Cauchy process (FGCP) with an additive and a multiplicative Gaussian white noise for describing subordinated anomalous fluctuations.  The FGCP displays intermittent dynamics during random time durations, whose analytical representation is given by the It$\hat{\rm o}$ stochastic integral.  The associated probability density function is given by the Tsallis $q$-Gaussian distribution at the stationary state. The method of fractional Feynman-Kac formula shows that weak ergodicity breaking of the FGCP depends on the existence of the subordinator and/or the divergence of variance.
%\begin{description}
%\end{description}
\end{abstract}

\pacs{Valid PACS appear here}% PACS, the Physics and Astronomy
\maketitle

%\tableofcontents

\section{\label{sec:sec1}Introduction}
Anomalous fluctuations associated with fat-tailed distributions have been observed in many complex systems.  In order to identify the anomalous fluctuations, Tsallis introduced the concept of nonextensive statistical mechanics (NESM), based on the principle of a generalized maximum entropy~\cite{Tsallis1}. Also, a nonlinear kinetic equation, known as a nonlinear Fokker-Planck equation (NFPE), provides the probability density function (PDF) for the NESM~\cite{Plastino}.  Applications of the NESM in various areas have been reported, such as, option pricing in stock markets~\cite{Tsallis2}, fully developed fluid turbulence~\cite{Arimitsu1}, magnetospheric self-organization~\cite{Pavlos1} and so on.  \par
As a dynamical foundation of the NESM, Beck introduced a stochastic differential equation (SDE) with a slowly fluctuating parameter~\cite{Beck1}.  His idea that the NESM results from inhomogeneous heat bath leads to the concept of superstatistics (SS)~\cite{Beck2, Beck3}.  On the other hand, the NFPE for a harmonic potential is equivalent to the Fokker-Planck equation (FPE) of a generalized Cauchy process (GCP) whereby the corresponding stochastic process is given by the solution of an SDE~\cite{Konno1}. The stationary PDF of the GCP provides the $q$-Gaussian distribution with a specific set of parameters.  \par
In the past few decades stochastic processes on random time durations have been extensively investigated, associated with continuous time random walks (CTRWs), where the evolution of the time variable is also a random variable following a waiting time distribution~\cite{Scher1,Metzler1,Klafter1}.  Under an appropriate scaling limit, a CTRW is approximated by a set of SDEs for coordinate and time variables~\cite{Fogedby1}.  In the case of power-law waiting time distributions, such SDE systems are alternatively expressed by subordinated stochastic differential equations (SSDEs), and then derive fractional Fokker-Planck equations (FFPEs) as their kinetic equations~\cite{Metzler2, Magdziarz1}.  \par
Related to comprehensive studies of the SSDE, their fluctuations are assumed to follow the Gaussian distribution.  However, we can observe anomalous fluctuations following non-Gaussian distributions in the real world: fluid particle transport in a rotating cylinder~\cite{Salomon}, a hopping cold atom in optical lattices~\cite{Lutz}, wave propagation in dissipative media~\cite{Uchiyama1}, etc.  For the sake of identifying the anomalous fluctuations on random time durations, the SSDE is a suitable model to describe non-Gaussian distributions.   \par
In this paper we introduce an SSDE as a model of anomalous fluctuations with the power-law waiting time distribution.  Section~\ref{sec:sec2} provides a brief review of the dynamical foundation of the NESM, where the concept of environmental fluctuations and a characteristic time scale are mentioned.  In Sec.~\ref{sec:sec3}, the basic idea of subordinated stochastic processes is presented and then is applied to the GCP.  Section~\ref{sec:sec4} shows the corresponding kinetic equation, where the analytical representation of the PDF and slow relaxation play a dominant role.  Section~\ref{sec:sec5} considers weak ergodicity breaking due to the existence of a subordinator or the divergence of the second moment, with the use of the fractional Feynman-Kac formula.  Section~\ref{sec:sec6} provides conclusions and aspects of future work.
\section{\label{sec:sec2}Dynamical foundation of nonextensive statistical mechanics}
The NESM is constructed based on Tsallis $q$-entropy, which is a generalization of the conventional Boltzmann-Gibbs entropy, being of the form
\begin{equation}
S_q[p]=\frac{1}{q-1}\left(1-\int[p(x)]^qdx\right),
\label{eq:QENT}
\end{equation}
where $p(x)$ is a PDF and $q$ is a positive real parameter.  Under the constraint for a potential function ${\Phi}(x)$,
%with respect to a state variable $x$,
%
\begin{equation}
\int{\Phi}(x)[p(x)]^qdx={\langle}{\Phi}(x){\rangle},
\label{eq:QCNST}
\end{equation}
the general form of the Tsallis distributions is derived as
\begin{equation}
p(x)=\frac{1}{Z_q[1+\tilde{\beta}(q-1){\Phi}(x)]^{\frac{1}{q-1}}},
\label{eq:QDIST}
\end{equation}
where $Z_p$ is a partition function defined by the normalization factor of $p(x)$, and $\tilde{\beta}$ is a positive real parameter.  In the case of harmonic potential, ${\Phi}(x)=x^2$, $p(x)$ in Eq.~(\ref{eq:QDIST}) leads to the $q$-Gaussian distribution.  \par
Alternatively, the NFPE has been developed as a macroscopic dynamical model of the NESM, namely, the kinetic equation of the PDF in Eq.~(\ref{eq:QDIST})~\cite{Plastino}.  The NFPE is given by
\begin{equation}
\frac{{\partial}f^{\mu}}{{\partial}t}=-\frac{{\partial}}{{\partial}x}[K(x,t)f^{\mu}]+\frac{{\partial}^2}{{\partial}x^2}[Q(x,t)f^{\nu}],
\label{eq:NFPE}
\end{equation}
where ${\mu}$ and ${\nu}$ are positive real parameters.  Although the NFPE in Eq.~(\ref{eq:NFPE}) is a nonlinear partial differential equation, analytical solutions can be obtained in specific cases~\cite{Plastino, Tsallis3}. \par
As a microscopic dynamical model of the NESM, whose statistical laws are derived from the $q$-Gaussian distribution, Beck introduced the Ornstein-Uhlenbeck (OU) process with fluctuating parameters:
\begin{equation}
dX=-{\gamma}Xdt+{\sigma}dW(t),
\label{eq:OU}
\end{equation}
where $W(t)$ is a Brownian motion, and positive real parameters ${\gamma}$ and ${\sigma}$ define a random parameter, known as inverse temperature, as ${\beta}={\gamma}/{\sigma}$, which follows the ${\chi}^2$ distribution with $n$ degrees of freedom as 
\begin{equation}
f({\beta})=\frac{1}{{\Gamma}\left(\frac{n}{2}\right)}\left(\frac{n}{2{\beta}_0}\right)^{\frac{n}{2}}{\beta}^{\frac{n}{2}-1}{\exp}\left(-\frac{n{\beta}}{2{\beta}_0}\right),
\label{eq:KI2}
\end{equation}
where ${\Gamma}(z)$ is the gamma function and ${\beta}_0$ is a positive real parameter.  It is assumed that the characteristic time scale of ${\beta}$ is much larger than that of $X(t)$ in Eq.~(\ref{eq:OU}).  In this situation, the conditional stationary PDF is obtained as
\begin{equation}
p(x|{\beta})=\sqrt{\frac{\beta}{2{\pi}}}{\exp}\left(-\frac{1}{2}{\beta}x^2\right).
\label{eq:GAUSSIAN}
\end{equation}
The marginalization with respect to ${\beta}$ for Eq.~(\ref{eq:GAUSSIAN}) with Eq.~(\ref{eq:KI2}), $p(x)={\int}p(x|{\beta})f({\beta})d{\beta}$, gives
\begin{equation}
p(x)=\frac{{\Gamma}\left(\frac{n+1}{2}\right)}{{\Gamma}\left(\frac{n}{2}\right)}\left(\frac{{\beta}_0}{{\pi}n}\right)^{\frac{1}{2}}\frac{1}{\left(1+\frac{{\beta}_0}{n}x^2\right)^{\frac{n+1}{2}}},
\label{eq:q-GAUSSIAN}
\end{equation}
which corresponds to the $q$-Gaussian distribution with the relations for $q$ and $\tilde{\beta}$ in Eq.~(\ref{eq:QDIST}): $q=1+2/(n+1)$ and $\tilde{\beta}=2{\beta}_0/(3-q)$.  \par
An alternative microscopic dynamical model of the NESM is the GCP driven by both additive and multiplicative noise in the form:
\begin{equation}
dX=-{\gamma}Xdt+X\sqrt{2D_m}dW_m(t)+\sqrt{2D_a}dW_a(t),
\label{eq:GCP}
\end{equation}
where ${\gamma}$ is a positive real parameter, $W_a(t)$ and $W_m(t)$ are Brownian motions with respective noise intensities $2D_a$ and $2D_m$~\cite{Konno1}.  The FPE of the GCP is the same form as the NFPE in Eq.~(\ref{eq:NFPE}) with ${\mu}=1$ in the case of a periodic potential.  The stationary PDF of the GCP, known as a generalized Cauchy distribution (GCD), is geven by
\begin{equation}
p(x)=\frac{a^{2b-1}}{B(b-1/2,1/2)}\frac{1}{(x^2+a^2)^b},
\label{eq:GCD}
\end{equation}
where $a=\sqrt{D_a/D_m}$, $b={\gamma}/2D_m$ and $B(z,z')$ is the beta function~\cite{Frank1}.  The GCD in Eq.~(\ref{eq:GCD}) is also the $q$-Gaussian distribution with relations that $q=b^{-1}$ and $\tilde{\beta}=a^{-2}b$ for Eq.~(\ref{eq:q-GAUSSIAN}).  In this case, the multiplicative noise in Eq.~(\ref{eq:GCP}) generates the power-law tails of the PDF, because without the term, Eq.~(\ref{eq:GCP}) reduces to the OU process.  
\section{\label{sec:sec3}Subordinated stochastic processes}
Random dynamics during random time durations are described by the CTRWs with waiting time distributions having a power-law tail.  A CTRW provides a kinetic equation, whose particle fluctuations obey a subordinated stochastic process.  In order to describe such fluctuations under an appropriate scaling limit, the SSDE is introduced in the form
\begin{equation}
Y(t)=X(S_t),
\label{eq:SSDE}
\end{equation}
where the subordinated stochastic process $Y(t)$ is generated from an SDE on intrinsic time ${\tau}$,
\begin{equation}
dX=F(X,{\tau})d{\tau}+G(X,{\tau})dW({\tau})
\label{eq:INTSDE}
\end{equation}
with $W({\tau})$ being a Brownian motion.  The subordinator $S_t$ in the right hand side of Eq.~(\ref{eq:SSDE}), which provides physical time evolution, is defined as
\begin{equation}
S_t={\rm inf}\{{\tau};U({\tau})>t\}
\label{eq:SOD}
\end{equation}
with $U({\tau})$ being a strictly increasing ${\alpha}$-stable L\'{e}vy motion, whose Laplace transform is given by ${\langle}e^{-kU({\tau})}{\rangle}=e^{{\tau}k^{\alpha}}$, where $0<{\alpha}<1$ \cite{Janicki1}.  \par
Based on the above notion of the SSDE, we construct a dynamical model for subordinated anomalous fluctuations.  The SDEs presented in Sec.~\ref{sec:sec2} are candidates for the intrinsic SDE of the subordinated dynamics.  For this purpose, however, the assumption of the first model in Sec.~\ref{sec:sec2} is contradictory because the characteristic time scale of the subordinated OU process diverges due to the power-law tail of the waiting time distribution.  On the other hand, the GCP has no assumption about the time scale, whereby it is employed as the intrinsic SDE of the SSDE, named subordinated GCP (SGCP).  We investigate statistical properties of the SGCP hereafter.  \par
To obtain a realization of the SGCP, it is necessary to solve the GCP with respect to intrinsic time ${\tau}$ in the form:
\begin{equation}
dX=-{\gamma}Xd{\tau}+X\sqrt{2D_m}dW_m({\tau})+\sqrt{2D_a}dW_a({\tau}),
\label{eq:ITGCP}
\end{equation}
where ${\gamma}$ is a positive real parameter, $W_a({\tau})$ and $W_m({\tau})$ are Brownian motions with respective noise intensities $2D_a$ and $2D_m$.  The GCP can be solved with the use of the It$\hat{\rm o}$ integral as
\begin{eqnarray}
&&X({\tau})=X(0)e^{-({\gamma}+D_m){\tau}+\sqrt{2D_m}W_m({\tau})}  \label{eq:SOL_ITGCP} \\
&&+\sqrt{2D_a}\int_0^te^{({\gamma}+D_m)({\tau}'-{\tau})-\sqrt{2D_m}[W_m({\tau}')-W_m({\tau})]}dW_a({\tau}') \nonumber 
\end{eqnarray}
with $X(0)$ being the initial value almost surely.  The combination of $X({\tau})$ in Eq.~(\ref{eq:SOL_ITGCP}) and the subordinator $S_t$ gives realizations of the SGCP.  \par
\begin{figure*}
\includegraphics{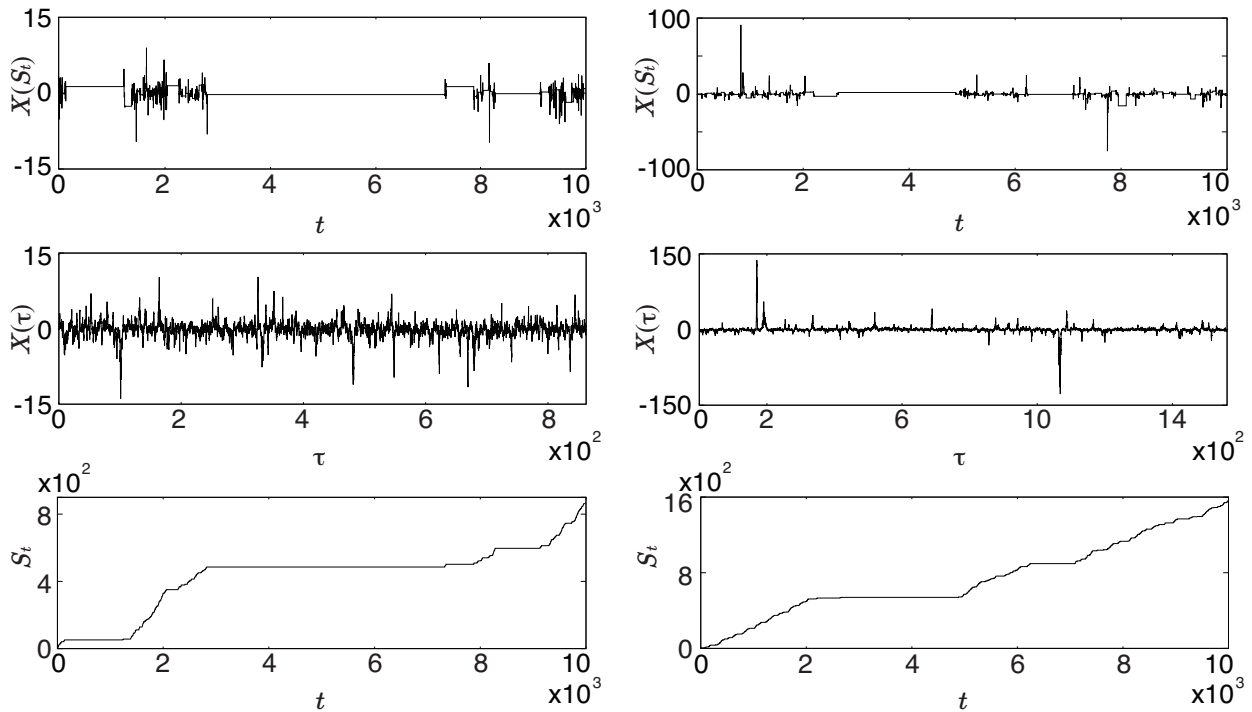}
\caption{\label{fig:realization}Realizations of the SGCP in the case of finite variance (left side: ${\gamma}=1.2$) and infinite variance (right side: ${\gamma}=0.7$).  The noise intensities and fractional index are fixed as $D_a=1.0, D_m=0.5$ and $ {\alpha}=0.7$.  The top figures are the realizations of the SGCP on physical time $t$.  The middle figures are the realizations of the GCP on intrinsic time $\tau$.  The bottom figures are the random processes of the subordinator $S_t$ defined by inversion of a positive one-sided L\'{e}vy process. }
\end{figure*}

Although the SGCP is solved analytically, it is often useful to carry out numerical simulations to obtain sample paths.  As a numerical integration scheme of Eq.~(\ref{eq:SOL_ITGCP}), a recursion relation of the GCP with respect to ${\tau}$ is obtained.  A discrete time ${\tau}_n=n{\Delta}{\tau}$, and random numbers ${\xi}_a$ and ${\xi}_m$, which follow mutually independent Gaussian distributions, gives a recursion relation of Eq.~(\ref{eq:SOL_ITGCP}) as
\begin{eqnarray}
X({\tau}_{n+1})&=&X({\tau}_n)e^{-({\gamma}+D_m){\Delta}{\tau}+\sqrt{2D_m}{\Delta}{\tau}{\xi}_m} \nonumber \\
&+&\sqrt{2D_a{\Delta}{\tau}}{\xi}_ae^{({\gamma}+D_m){\Delta}{\tau}-\sqrt{2D_m}{\Delta}{\tau}{\xi}_m}.
\label{eq:DIS_INTGCP}
\end{eqnarray}
The numerical scheme for solving SSDEs presented in Ref.~\cite{Magdziarz1} provides realizations of the SGCP as shown in Fig.~\ref{fig:realization}.  The left and right figures are realizations of the SGCP in the case of finite and infinite variances, respectively.  It is seen that the magnitude of these realizations shows fluctuations of different orders depending on the value of the parameters.  The realizations of the SGCP on physical time (top figures) are generated from those on intrinsic time (middle figures) and the corresponding subordinators (bottom figures).  It is seen that intermittent dynamics are under the constraint of long time durations due to the subordinator.  Thus the SGCP can describe the large fluctuations in magnitude with long memory, which are often investigated in complex systems.
\par
\section{\label{sec:sec4}Fractional Fokker-Planck equation}
\begin{figure}
\includegraphics{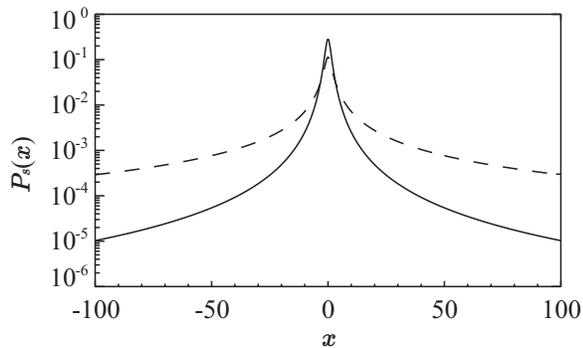}
\caption{\label{fig:PDF}The GCDs derived from Eqs.~(\ref{eq:FFPE}) and (\ref{eq:LFP}) on stationary states.  The solid line shows the case of finite variance with ${\gamma}=1.2$, $D_a=1.0$ and$D_m=0.5$.  The dashed line shows the case of infinite variance with ${\gamma}=0.7$, $D_a=1.0$ and $D_m=0.5$, corresponding to the realizations in Fig.~\ref{fig:realization}.  It is shown that the tails of the latter case are broader than those of the former case. }
\end{figure}

Ensemble dynamics of the subordinated stochastic processes determine their kinetics.  The evolution equation for the PDF of the SSDE is generally given as a fractional Fokker-Planck equation (FFPE), which consists of the Fokker-Planck operator $L_{FP}$ derived from the SDE in Eq.~(\ref{eq:INTSDE}) and the Riemann-Liouville fractional derivative ${}_0{\bf D}^{1-{\alpha}}_t$~\cite{Podlubny1}, in the form: 
\begin{equation}
\frac{\partial}{{\partial}{t}}P(x,t)={}_0{\bf D}^{1-{\alpha}}_tL_{FP}P(x,t),
\label{eq:FFPE}
\end{equation}
where the Riemann-Liouville fractional derivative for a time dependent function is defined by
\begin{equation}
{}_0{\bf D}^{1-{\alpha}}_tf(t)=\frac{1}{\Gamma({\alpha})}\frac{d}{dt}\int_0^t(t-{t'})^{{\alpha}-1}f(t')dt'
\label{eq:RLDT}
\end{equation}
with ${\Gamma}(z)$ being the gamma function~\cite{Frank1}.  Here, in the case of the SGCP, the corresponding Fokker-Planck operator is given by
\begin{equation}
L_{FP}=\frac{\partial}{{\partial}x}{\gamma}x+\frac{{\partial}^2}{{\partial}x^2}(D_mx^2+D_a),
\label{eq:LFP}
\end{equation}
which is the same form as that of the GCP~\cite{Konno1}.  On equilibrium state, the GCD is derived from $L_{FP}$ in Eq.~(\ref{eq:LFP}) as the stationary PDF in the form:  
\begin{equation}
P_s(x)=\frac{a^{2b-1}}{B(b-1/2,1/2)}\frac{1}{(x^2+a^2)^b}.
\label{eq:GCD2}
\end{equation}
The value of the parameter $b={\gamma}/2D_m$ determines whether the variance is finite or not, as is shown in Fig.~\ref{fig:PDF}: The solid line shows the case of finite variance, ${\gamma}>2D_m$, the dashed line shows that of infinite variance, ${\gamma}<2D_m$.  The respective cases correspond to the realizations of the SGCP in Fig.~\ref{fig:realization}, which are at a quite different order of magnitude.  \par
The transition PDF of the SGCP is obtained by solving the FFPE in Eq.~(\ref{eq:FFPE}) with the Fokker-Planck operator in Eq.~(\ref{eq:LFP}).  The inverse L\'{e}vy transform (ILT) is utilized to solve the FFPE~\cite{Klafter1}.  For the solution of the FFPE, $P(x,t)$, the ILT is introduced as 
\begin{equation}
P(x,t)=\int_0^{\infty}p(x,{\tau})T({\tau},t)d{\tau},
\label{eq:ILTF}
\end{equation}
where the kernel function $T({\tau},t)$ is given in the Laplace space as
\begin{equation}
T({\tau},s)=s^{{\alpha}-1}e^{-{\tau}s^{\alpha}}
\label{eq:IKLS}
\end{equation}
and $p(x,{\tau})$ is the solution of the conventional Fokker-Planck equation (FPE) with $L_{FP}$ in Eq.~(\ref{eq:LFP}) as
\begin{equation}
\frac{\partial}{{\partial}{\tau}}p(x,{\tau})=L_{FP}p(x,{\tau}).
\label{eq:FPE}
\end{equation}
In general the FPE is solved by the method of eigenfunction expansion.  In this case, $L_{FP}$ in Eq.~(\ref{eq:LFP}) has both discrete and continuous eigenvalues, and thus gives the expression of $p(x,{\tau})$ as
%
% for preprint
%\begin{eqnarray}
%p(x,{\tau}|x_0,{\tau}_0)&=&e^{U(x)/2-U(x_0)/2}\Biggl[\sum_{n=0}^N{\phi}_n(x){\phi}_n(x_0)e^{-{\lambda}_n({\tau}-{\tau}_0)} \nonumber \\
%&+&\int{\phi}_{\lambda}(x){\phi}_{\lambda}(x_0)e^{-{\lambda}({\tau}-{\tau}_0)}d{\lambda}\Biggr],
%\label{eq:EFEP}
%\end{eqnarray}
%
% for submit
\begin{widetext}
\begin{equation}
p(x,{\tau}|x_0,{\tau}_0)=e^{U(x)/2-U(x_0)/2}\Biggl[\sum_{n=0}^{\infty}{\phi}_n(x){\phi}_n(x_0)e^{-{\lambda}_n{\tau}}+\int{\phi}_{\lambda}(x){\phi}_{\lambda}(x_0)e^{-{\lambda}{\tau}}d{\lambda}\Biggr],
\label{eq:EFEP}
\end{equation}
\end{widetext}
where $\{{\lambda}_n,{\phi}_n({\cdot})\}$ and $\{{\lambda},{\phi}({\cdot})\}$ are pairs of eigenvalues and eigenfunctions for the discrete and continuous cases, respectively.  The specific form of the eigenfunction expansion in Eq.~(\ref{eq:EFEP}) is given by the hypergeometric function as explained in Appendix~\ref{app:appA}.  Note that the discrete eigenvalues vanish when ${\gamma}<2D_m$, namely, the variance of the GCD in Eq.~(\ref{eq:GCD2}) is infinite.  \par
The ILT in Eq.~(\ref{eq:ILTF}) converts the eigenfunction expansion in Eq.~(\ref{eq:EFEP}) to that of the FFPE in Eq.~(\ref{eq:FFPE}) as
\begin{widetext}
\begin{equation}
P(x,t|x_0,t_0)=e^{U(x)/2-U(x_0)/2}
\left[\sum_{n=0}^N{\phi}_n(x){\phi}_n(x_0)E_{{\alpha},1}[-{\lambda}_n(t-t_0)^{\alpha}]+\int{\phi}_{\lambda}(x){\phi}_{\lambda}(x_0)E_{{\alpha},1}[-{\lambda}(t-t_0)^{\alpha}]d{\lambda}\right],
\label{eq:EFEP2}
\end{equation}
\end{widetext}
where $E_{{\alpha},{\beta}}(t)$ is the generalized Mittag-Leffler function defined by
\begin{equation}
E_{{\alpha},{\beta}}(t)=\sum_{n=0}^{\infty}\frac{t^n}{\Gamma({\alpha}n+{\beta})}
\label{eq:MLF}
\end{equation}
with ${\Gamma}(z)$ being the gamma function.  Eventually, statistics of the SGCP are calculated with the transition PDF in Eq.~(\ref{eq:EFEP2}).  \par
\begin{figure}
\includegraphics{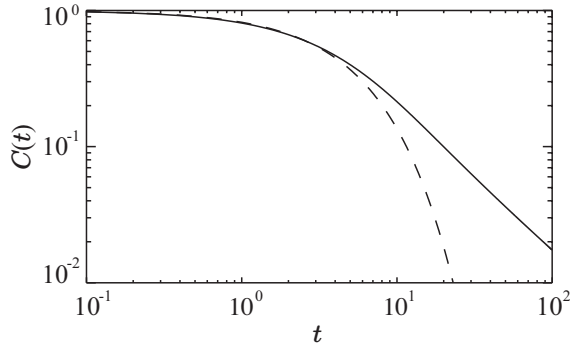}
\caption{\label{fig:ACF}ACF of the SGCP (solid line) and the GCP (dashed line) with the parameters ${\gamma}=0.7$, $D_m=0.5$ and ${\alpha}=0.7$.  Relaxation of the SGCP is slower than that of the GCP because of the power-law tail.  }
\end{figure}

Alternatively, the ILT in Eq.~(\ref{eq:ILTF}) gives the formula for calculating statistics of a random variable generated from FFPEs as
\begin{equation}
{\langle}f(X(t)){\rangle}=\int{\langle}f(X({\tau})){\rangle}T({\tau},t)d{\tau}
\label{eq:ALTST}
\end{equation}
with $f(X)$ being a function of a random variable $X$.  Since the normalized autocorrelation function (ACF) for the GCP is derived from Eq.~(\ref{eq:SOL_ITGCP}) as $C({\tau})=e^{-({\gamma}-D_m){\tau}}$ and the ILT of the exponential function is the Mittag-Leffler function, the ACF for the SGCP is given as
\begin{equation}
C(t)=E_{{\alpha},1}[-({\gamma}-D_m)t^{\alpha}].
\label{eq:ACF}
\end{equation}
In Fig.~(\ref{fig:ACF}), the ACF for the SGCD displays slow relaxation due to the long time durations generated by the subordinator. \par
As is shown in Eqs.~(\ref{eq:GCD}) and~(\ref{eq:ACF}), the SGCP provides the fat-tailed PDF and the ACF with a long time tail, which respectively correspond to large deviations and long memory of fluctuations.  Therefore, in many complex systems. the SGCP can be utilized to describe anomalous fluctuations with a subordinator.
\section{\label{sec:sec5}Weak ergodicity breaking}
Ergodicity is one of the most fundamental properties of statistical mechanics, which guarantees the equivalence between the ensemble average and the time average in the equilibrium state.  Based on ergodicity, thus, one can implement a statistical analysis of experimental data or a time series analysis using the tools for time averaging.  To investigate ergodicity, Khinchin's theorem, which shows that an ACF converging to zero guarantees ergodicity, was developed~\cite{Khinchin1}.  Nevertheless, it was proved that the subordinated OU process, whose ACF converges to zero, shows weak ergodicity breaking~\cite{Turgeman1}.  Here, we investigate whether ergodicity of the SGCP breaks or not by means of a kinetic equation for a Brownian functional.  \par
The Brownian functional of a stochastic process $\{X(t)\}_{t>0}$ is defined by $A(t)=\int_0^tU[X({\tau})]d{\tau}$ with a prescribed function $U(x)$.  Then time average is given by $\bar{X}(t)=A(t)/t$ with $U(x)=x$.  When the PDF of $\bar{X}(t)$ leads to Dirac delta function, namely, the variance of $\bar{X}(t)$ converges to zero, at infinite time limit, $\{X(t)\}_{t>0}$ is confirmed to be ergodic.  It is thus necessary to obtain the PDF for $A(t)$ by calculating the statistics of the Brownian functional.  \par
To calculate the Brownian functional of a subordinated stochastic process, the fractional Feynman-Kac equation (FFKE) with a fractional substantial derivative was introduced~\cite{Turgeman1}.  In the case of the SGCP, the corresponding FFKE is of the form:
\begin{equation}
\frac{\partial}{{\partial}t}G(x,A,t)=L_{FP}{\cal D}_t^{1-{\alpha}}G(x,A,t)-x\frac{\partial}{{\partial}A}G(x,A,t),
\label{eq:FFKE}
\end{equation}
where the Fokker-Planck operator is the same as $L_{FP}$ in Eq.~(\ref{eq:LFP}) and the fractional substantial derivative is defined by $\mathcal{D}_t^{1-{\alpha}}=({\partial}/{{\partial}t}+x{\partial}/{{\partial}A})^{1-{\alpha}}$ with a real parameter $0<{\alpha}<1$.  With the help of the Laplace transform, $\tilde{G}(x,A,s)=\int_0^{\infty}G(x,A,t)e^{-st}dt$, the variance of $A$ can be obtained in the Laplace space as
\begin{equation}
{\langle}(A-{\langle}A{\rangle})^2{\rangle}=\frac{2}{s^3}\frac{(1-{\alpha}){\gamma}+s^{\alpha}}{{\gamma}+s^{\alpha}}\frac{2D_a+x_0^2s^{\alpha}}{2({\gamma}-D_m)+s^{\alpha}}
\label{eq:VRA}
\end{equation}
with $x_0$ being the initial value of $X(t)$ almost surely.  Derivation of the moments with respect to $X$ and $A$ up to the second-order are shown in Appendix~\ref{app:appB}.  With the use of the Laplace transform of the Mittag-Leffler function, $\int_0^{\infty}e^{-st}E_{{\alpha},3}(-ct^{\alpha})dt=s^{{\alpha}-3}/(s^{\alpha}+c)$, ${\langle}\bar{X}^2{\rangle}$ in the time domain is obtained as
\begin{eqnarray}
{\langle}\bar{X}^2{\rangle}&=&\frac{(1-{\alpha})D_a}{{\gamma}-D_m}+\frac{2{\alpha}(2D_a-x_0^2)}{{\gamma}-2D_m}E_{{\alpha},3}(-{\gamma}t^{\alpha}) \nonumber \\ 
&+&\frac{2[x_0^2({\gamma}-D_m)({\gamma}+{\alpha}-2D_m)-(1-{\alpha}){\gamma}+2D_m]}{({\gamma}-D_m)({\gamma}-2D_m)} \nonumber \\
&{\times}&E_{{\alpha},3}(-2({\gamma}-D_m)t^{\alpha}).
\label{eq:MSD}
\end{eqnarray}
When ${\gamma}>D_m$, the infinite time limit of ${\langle}\bar{X}^2{\rangle}$ leads to a finite value as $\lim_{t{\to}{\infty}}{\langle}\bar{X}^2{\rangle}=(1-{\alpha})D_a/({\gamma}-D_m)$, except ${\alpha}=1$.  On the other hand, when ${\gamma}{\leq}D_m$,  $\lim_{t{\to}{\infty}}{\langle}\bar{X}^2{\rangle}={\infty}$.  This result means that the case of the GCP with finite variance only guarantees ergodicity whereas the GCP with infinite variance and the SGCP with both finite and infinite variance show weak ergodicity breaking.  In other words, one should carefully choose  whether to use an ensemble or a time average to implement a statistical analysis for empirical intermittent fluctuations with random time durations.
\section{\label{sec:sec6}Conclusions}
We have reviewed the two alternative models for dynamical foundation of the NESM, where the $q$-Gaussian distribution, as the specific case of  the GCD, is given at the stationary state.  In both cases the environmental fluctuations play the dominant role in generating the power-law tails of the PDF.  \par
To employ these models as the intrinsic SDE of the subordinated stochastic process, we have considered the characteristic time scales in the fluctuating environment, and then have pointed out that the SGCP is the suitable model for describing the anomalous fluctuations with the subordinator.  The statistical properties of the SGCP have been investigated with the use of the FFPE.  The method of eigenfunction expansion and the ILT provide an analytical description of the transition PDF, which is available to calculate the moments of the SGCP, expressed by the combination of the Mittag-Leffler function and the hypergeometric function.  \par
Weak ergodicity breaking of the SGCP is ascribed to the existence of the subordinator or infinite variance by means of the FFKE.  Thus, statistical analysis for subordinated intermittent stochastic processes should be implemented only by ensemble averaging to obtain correct estimates. \par
The alternative GCP with a logarithmic potential also follows the GCD at the stationary state \cite{Lutz}.  The eigenfunction expansion of this GCP is different from that of the GCP handled in this paper.  Specific analysis of the difference between the GCPs will be our future work.
% \begin{acknowledgments}
% \end{acknowledgments}

\appendix

\section{\label{app:appA}Eigenfunction expansion of the generalized Cauchy process}
The GCP in Eq.~(\ref{eq:ITGCP}) is transformed into the canonical form by replacing ${\gamma}/(2D_m){\to}{\gamma}-1/2$ and $D_m{\tau}{\to}{\tau}$ as 
\begin{equation}
dX=(1-2{\gamma})Xd{\tau}+\sqrt{2(X^2+a^2)}dW(\tau).
\label{eq:CGCP_APP}
\end{equation}
By change of the variable, $X=a\sinh{Y}$, Eq.~(\ref{eq:CGCP_APP}) leads to
\begin{equation}
dY=-\frac{dU}{dY}d{\tau}+\sqrt{2}dW(\tau)
\label{eq:CGCP2_APP}
\end{equation}
with the potential function $U(Y)=2{\gamma}\ln(\cosh{Y})$.  The corresponding FPE is readily obtained as 
\begin{equation}
\frac{\partial}{{\partial}{\tau}}p(y,{\tau})=\frac{\partial}{{\partial}y}\left[\frac{dU}{dy}+\frac{\partial}{{\partial}y}\right]p(y,{\tau}).
\label{eq:FPE_APP}
\end{equation}
The unitary transform, $p(y,{\tau})=e^{-U(y)/2-{\gamma}^2{\tau}}{\psi}(y,{\tau})$, gives the imaginary time Schr\"{o}dinger equation as
\begin{equation}
-\frac{\partial}{{\partial}{\tau}}{\psi}(y,{\tau})=\left[-\frac{{\partial}^2}{{\partial}y^2}+V(y)\right]{\psi}(y,{\tau}),
\label{eq:ITSE_APP}
\end{equation}
where the potential function $V(y)$ is given by
\begin{equation}
V(y)=-{\gamma}({\gamma}+1){\rm sech}^2y.
\label{eq:PTNT_APP}
\end{equation}
The form of $V(y)$ in Eq.~(\ref{eq:PTNT_APP}) is known as the modified P\"{o}schel type potential, which has, depending on the parameter ${\gamma}$, discrete and continuous eigenvalues since $\lim_{y{\to}{\pm}{\infty}}V(y)=0$~\cite{Fluegge1}.  Let us take the separation of variables as ${\psi}(y,{\tau})=e^{-{\lambda}{\tau}}{\phi}(y)$ with ${\lambda}$ being the eigenvalue.  For a continuous eigenvalue, ${\lambda}=k^2$ gives the time-independent Schr\"{o}dinger equation as
\begin{equation}
\frac{d^2{\phi}}{dy^2}+\left[k^2+{\gamma}({\gamma}+1){\rm sech}^2y\right]{\phi}=0.
\label{eq:TISE_APP}
\end{equation}
Changing the variable, $z={\cosh}^2y$, and taking the ansatz, ${\phi}(y)=z^{({\gamma}+1)/2}{\varphi}(z)$, Eq.~(\ref{eq:TISE_APP}) yields the hypergeometric differential equation,
\begin{eqnarray}
&&z(z-1)\frac{d^2{\varphi}}{dz^2}+\left[\left({\gamma}+\frac{3}{2}\right)-({\gamma}+2)z\right]\frac{d{\varphi}}{dz} \nonumber \\
&&-\frac{({\gamma}+1)^2+k^2}{4}{\varphi}=0.
\label{eq:HGDE_APP}
\end{eqnarray}
Thus ${\varphi}(z)$ is expressed by superposition of the hypergeometric functions, $F({\xi},{\eta};{\zeta};z)$~\cite{Frank1}, as
\begin{eqnarray}
{\varphi}(z)&=&AF\left({\alpha},{\beta};\frac{1}{2};1-z\right) \nonumber \\
&+&B(1-z)^{\frac{1}{2}}F\left({\alpha}+\frac{1}{2},{\beta}+\frac{1}{2};\frac{3}{2};1-z\right),
\label{eq:SOLHGDE_APP}
\end{eqnarray}
where the coefficients $A$ and $B$ are determined by boundary conditions, and the parameters ${\alpha}$ and ${\beta}$ are defined by ${\alpha}=({\gamma}+1+ik)/2$ and ${\beta}=({\gamma}+1-ik)/2$.  Assume that ${\phi}(y)$ consists of an even function ${\phi}_e(y)$ and an odd function ${\phi}_o(y)$, then ${\phi}(y)$ is expressed by ${\phi}(y)=A{\phi}_e(y)+B{\phi}_o(y)$ with 
\begin{equation}
{\phi}_e(y)=(\cosh{y})^{{\gamma}+1}F\left({\alpha},{\beta};\frac{1}{2};-{\sinh}^2y\right)
\label{eq:EVEN_APP}
\end{equation}
and
\begin{equation}
{\phi}_o(y)=(\cosh{y})^{{\gamma}+1}F\left({\alpha}+\frac{1}{2},{\beta}+\frac{1}{2};\frac{3}{2};-{\sinh}^2y\right).
\label{eq:ODD_APP}
\end{equation}
A well-known formula of the hypergeometric function,
%
% for preprint
%\begin{eqnarray}
%&&\frac{F({\kappa},{\mu};{\nu};{\zeta})}{{\Gamma}({\nu})}
%=
%\frac{(-{\zeta})^{-{\kappa}}{\Gamma}({\mu}-{\kappa})}{{\Gamma}({\nu}-{\kappa}){\Gamma}({\mu})}F\left({\kappa},{\kappa}-{\nu}+1;{\kappa}-{\mu}+1;\frac{1}{\zeta}\right) \nonumber \\
%&&+
%\frac{(-{\zeta})^{-{\mu}}{\Gamma}({\kappa}-{\mu})}{{\Gamma}({\nu}-{\mu}){\Gamma}({\kappa})}F\left({\kappa},{\mu}-{\nu}+1;{\mu}-{\kappa}+1;\frac{1}{\zeta}\right),
%\label{eq:HGFF_APP}
%\end{eqnarray}
%
% for submit
\begin{widetext}
\begin{equation}
\frac{F({\kappa},{\mu};{\nu};{\zeta})}{{\Gamma}({\nu})}
=
\frac{(-{\zeta})^{-{\kappa}}{\Gamma}({\mu}-{\kappa})}{{\Gamma}({\nu}-{\kappa}){\Gamma}({\mu})}F\left({\kappa},{\kappa}-{\nu}+1;{\kappa}-{\mu}+1;\frac{1}{\zeta}\right)
+
\frac{(-{\zeta})^{-{\mu}}{\Gamma}({\kappa}-{\mu})}{{\Gamma}({\nu}-{\mu}){\Gamma}({\kappa})}F\left({\kappa},{\mu}-{\nu}+1;{\mu}-{\kappa}+1;\frac{1}{\zeta}\right),
\label{eq:HGFF_APP}
\end{equation}
\end{widetext}
and the asymptotic form of ${\sinh}^2y$ in Eqs.~(\ref{eq:EVEN_APP}) and (\ref{eq:ODD_APP}) give, on $|y|{\to}{\infty}$, 
%
%\begin{widetext}
%\begin{equation}
%\lim_{|y|{\to}{\infty}}{\varphi}_e(y)
%=
%{\Gamma}\left(\frac{1}{2}\right)\left[\frac{{\Gamma}(-ik)e^{-ik\log{2}}}{{\Gamma}\left(\frac{{\gamma}+1}{2}-i\frac{k}{2}\right){\Gamma}\left(-\frac{\gamma}{2}-i\frac{k}{2}\right)}e^{-k|y|}
%+
%\frac{{\Gamma}(ik)e^{-ik\log{2}}}{{\Gamma}\left(\frac{{\gamma}+1}{2}+i\frac{k}{2}\right){\Gamma}\left(-\frac{\gamma}{2}+i\frac{k}{2}\right)}e^{k|y|}\right],
%\label{eq:ASMEVEN_APP}
%\end{equation}
%\end{widetext}
%
%
\begin{eqnarray}
{\phi}_e(y)
&{\to}&
{\Gamma}\left(\frac{1}{2}\right)\Biggl[\frac{{\Gamma}(-ik)e^{ik\log{2}}}{{\Gamma}\left(\frac{{\gamma}+1}{2}-i\frac{k}{2}\right){\Gamma}\left(-\frac{\gamma}{2}-i\frac{k}{2}\right)}e^{-k|y|} \nonumber \\
&+&
\frac{{\Gamma}(ik)e^{-ik\log{2}}}{{\Gamma}\left(\frac{{\gamma}+1}{2}+i\frac{k}{2}\right){\Gamma}\left(-\frac{\gamma}{2}+i\frac{k}{2}\right)}e^{k|y|}\Biggr],
\label{eq:ASMEVEN_APP}
\end{eqnarray}
and,
%
%\begin{widetext}
%\begin{equation}
%\lim_{|y|{\to}{\infty}}{\varphi}_o(y)
%=
{%\rm sgn}(y){\Gamma}\left(\frac{3}{2}\right)\left[\frac{{\Gamma}(-ik)e^{-ik\log{2}}}{{\Gamma}\left(\frac{{\gamma}+2}{2}-i\frac{k}{2}\right){\Gamma}\left(-\frac{{\gamma}+1}{2}-i\frac{k}{2}\right)}e^{-k|y|}
%+
%\frac{{\Gamma}(ik)e^{-ik\log{2}}}{{\Gamma}\left(\frac{{\gamma}+2}{2}+i\frac{k}{2}\right){\Gamma}\left(-\frac{{\gamma}+1}{2}+i\frac{k}{2}\right)}e^{k|y|}\right],
%\label{eq:ASMODD_APP}
%\end{equation}
%\end{widetext}
%
%
\begin{eqnarray}
{\phi}_o(y)
&{\to}&
{\pm}{\Gamma}\left(\frac{3}{2}\right)\Biggl[\frac{{\Gamma}(-ik)e^{ik\log{2}}}{{\Gamma}\left(\frac{{\gamma}+2}{2}-i\frac{k}{2}\right){\Gamma}\left(-\frac{{\gamma}+1}{2}-i\frac{k}{2}\right)}e^{-k|y|} \nonumber \\
&+&
\frac{{\Gamma}(ik)e^{-ik\log{2}}}{{\Gamma}\left(\frac{{\gamma}+2}{2}+i\frac{k}{2}\right){\Gamma}\left(-\frac{{\gamma}-1}{2}+i\frac{k}{2}\right)}e^{k|y|}\Biggr],
\label{eq:ASMODD_APP}
\end{eqnarray}
where ${\pm}$ in the right hand side of Eq.~(\ref{eq:ASMODD_APP}) depends on the sign of $y$.  The coefficients $A$ and $B$ are determined on scattering states,
\begin{equation}
  {\phi}(y) = \begin{cases}
    e^{iky}+Re^{-iky} & (y{\to}-{\infty}) \\
    Te^{iky} & (y{\to}+{\infty})
  \end{cases}
\label{eq:SCT_APP}
\end{equation}
with $|T|^2={\sin}^2({\theta}_e-{\theta}_o)$ and $|R|^2={\cos}^2({\theta}_e-{\theta}_o)$, as
\begin{equation}
A=\frac{e^{i{\theta}_e}}{2{\Gamma}\left(\frac{1}{2}\right)r_e},\;\;r_ee^{i{\theta}_e}=\frac{{\Gamma}(ik)e^{-ik\log{2}}}{{\Gamma}\left(\frac{{\gamma}+1}{2}+i\frac{k}{2}\right){\Gamma}\left(-\frac{\gamma}{2}+i\frac{k}{2}\right)},
\label{eq:A_APP}
\end{equation}
and
\begin{equation}
B=-\frac{e^{i{\theta}_o}}{2{\Gamma}\left(\frac{3}{2}\right)r_o},\;r_oe^{i{\theta}_o}=\frac{{\Gamma}(ik)e^{-ik\log{2}}}{{\Gamma}\left(\frac{{\gamma}+2}{2}+i\frac{k}{2}\right){\Gamma}\left(-\frac{{\gamma}-1}{2}+i\frac{k}{2}\right)}.
\label{eq:B_APP}
\end{equation}
\par
Discrete eigenvalues are estimated by setting ${\lambda}=-l^2$.  In this case the parameters ${\alpha}$ and ${\beta}$ in Eq.~(\ref{eq:SOLHGDE_APP}) are replaced by ${\alpha}=({\gamma}+1-l)/2$ and ${\beta}=({\gamma}+1+l)/2$.  To avoid ${\phi}(y)$ diverging, the coefficients of $e^{l|y|}$ should be zero for both ${\phi}_e(y)$ and ${\phi}_o(y)$.  Hence discrete eigenvalues are readily obtained as 
\begin{equation}
  l = \begin{cases}
    {\gamma}-2n & {\rm for}\;\;{\phi}_e(y) \\
    {\gamma}-2n-1 & {\rm for}\;\;{\phi}_o(y)
  \end{cases}
\label{eq:DISEV_APP}
\end{equation}
with $n$ being the natural number, which corresponds to the respective even and odd functions.  Note that the discrete eigenvalue can exist when ${\gamma}>1$.  That situation corresponds to the case of finite variance.  
%Thus $n$-dependent eigenfunction is described as ${\varphi}_n(y)=A_n{\varphi}_{e,n}(y)+B_n{\varphi}_{o,n}(y)$ with respective even and add functions
%
%\begin{equation}
%{\varphi}_{e,n}(y)={\Gamma}\left(\frac{1}{2}\right)\frac{{\Gamma}(2n-{\gamma})e^{({\gamma}-2n){\log}2}}{{\Gamma}\left(\frac{2n+1}{2}\right){\Gamma}(n-{\gamma})}e^{-({\gamma}-2n)|y|}
%\label{eq:NEFEVEN_APP}
%\end{equation}
%
%and
%
%\begin{equation}
%{\varphi}_{o,n}(y)={\pm}{\Gamma}\left(\frac{3}{2}\right)\frac{{\Gamma}(2n+1-{\gamma})e^{({\gamma}-2n-1){\log}2}}{{\Gamma}\left(\frac{2n+3}{2}\right){\Gamma}(n+1-{\gamma})}e^{-({\gamma}-2n-1)|y|}.
%\label{eq:NEFEVEN_APP}
%\end{equation}
%
\par
Eventually, the eigenfunction expansion of ${\psi}(y,{\tau})$ in Eq.~(\ref{eq:ITSE_APP}) is completed as
%
%for preprint
%\begin{eqnarray}
%{\psi}(y,{\tau})&=&\sum_{n=0}^{n<\frac{\gamma}{2}}\left[A_n{\phi}_{e,n}(y)e^{-({\gamma}-2n)^2{\tau}}+B_n{\phi}_{o,n}(y)e^{-({\gamma}-2n-1)^2{\tau}}\right] \nonumber \\
%&+&
%2\int_0^{\infty}k\left[A{\phi}_e(y)+B{\phi}_o(y)\right]e^{-k^2{\tau}}dk.
%\label{eq:EFEXP_APP}
%\end{eqnarray}
%
% for submit
\begin{widetext}
\begin{equation}
{\psi}(y,{\tau})=\sum_{n=0}^{n<\frac{\gamma}{2}}\left[A_n{\phi}_{e,n}(y)e^{-({\gamma}-2n)^2{\tau}}+B_n{\phi}_{o,n}(y)e^{-({\gamma}-2n-1)^2{\tau}}\right]+2\int_0^{\infty}k\left[A{\phi}_e(y)+B{\phi}_o(y)\right]e^{-k^2{\tau}}dk.
\label{eq:EFEXP_APP}
\end{equation}
\end{widetext}
\section{\label{app:appB}Moments of the fractional Feynman-Kac equation}
The FFKE in the Laplace space is of the form:
\begin{eqnarray}
s\hat{G}(x,A,s)-G_0(x,A)&=&L_{FP}{\cal D}_s^{1-{\alpha}}\hat{G}(x,A,s) \nonumber \\
&-&x\frac{\partial}{{\partial}A}\hat{G}(x,A,s).
\label{eq:LTFFKE_APP}
\end{eqnarray}
In the Laplace space the fractional substantial derivative, ${\cal D}_s^{1-{\alpha}}=(s+x{\partial}/{\partial}A)^{1-{\alpha}}$, can be expanded as
\begin{equation}
{\cal D}_s^{1-{\alpha}}=s^{1-{\alpha}}\left(1+s^{-1}x\frac{\partial}{{\partial}A}\right)\sum_{r=0}^{\infty}\left(s^{-1}x\frac{\partial}{{\partial}A}\right)^r.
\label{eq:FSDEXP_APP}
\end{equation}
The expansion of ${\cal D}_s^{1-{\alpha}}$ is terminated at most at the second order in the case of the FGCP when calculating moments with respect to $X$ and $A$.  Hence all of the moments are readily obtained as
\begin{eqnarray}
&&{\langle}x(s){\rangle}=\frac{s^{{\alpha}-1}}{s^{\alpha}+{\gamma}}x_0, \\
&&{\langle}A(s){\rangle}=\frac{1}{s}{\langle}x(s){\rangle}, \\
&&{\langle}x^2(s){\rangle}=\frac{1}{s}\frac{2D_a+x_0^2s^{\alpha}}{s^{\alpha}+2({\gamma}-D_m)}, \\
&&{\langle}x(s)A(s){\rangle}=\frac{1}{s}\frac{(1-{\alpha}){\gamma}+s^{\alpha}}{s^{\alpha}+{\gamma}}{\langle}x^2(s){\rangle}, \\
&&{\langle}A^2(s){\rangle}=\frac{2}{s}{\langle}x(s)A(s){\rangle}.
\label{eq:MOMENTS_APP}
\end{eqnarray}
It is shown that the convergence of the second order moments depends on the balance between the dissipation coefficient ${\gamma}$ and the multiplicative noise intensity $D_m$.

\nocite{*}

\bibliography{PRE2_references.bbl}% Produces the bibliography via BibTeX.

\end{document}